\documentclass[useAMS,usenatbib]{mn2e}

\usepackage[dvips]{graphicx}

\title[Monte-Carlo simulations of the broadband X-ray continuum of SS433]{Monte-Carlo simulations of the broadband X-ray continuum of SS433}
\author[Yu. M. Krivosheyev, G. S. Bisnovatyi-Kogan, A. M. Cherepashchuk and K. A. Postnov]{Yu. M. Krivosheyev$^{1,4}$\thanks{E-mail:
krivosheev@iki.rssi.ru (YuMK); gkogan@iki.rssi.ru (GSBK);
cher@sai.msi.ru (AMCh); pk@sai.msi.ru (KAP)}, G. S.
Bisnovatyi-Kogan$^{1,3,4}$, \medskip\\ \rm \LARGE A. M.
Cherepashchuk$^{2}$ and  K. A. Postnov$^{2}$\\
$^{1}$Space Research Institute of Russian Academy of Science (IKI), Profsoyuznaya 84/32, Moscow 117997, Russia\\
$^{2}$Sternberg Astronomical Institute, Moscow State University, Universitetsky pr., 13, Moscow 119992, Russia\\
$^{3}$Joint Institute for Nuclear Research, Joliot-Curie 6, Dubna 141980, Moscow region, Russia\\
$^{4}$Moscow Engineering Physics Institute, Kashirskoe Shosse 31, Moscow 115409, Russia}
\begin{document}

\pagerange{\pageref{firstpage}--\pageref{lastpage}} \pubyear{2008}

\maketitle

\label{firstpage}

\begin{abstract}
We develop a Monte-Carlo technique based on L.B. Lucy's indivisible photon packets method to calculate
X-ray continuum spectra of comptonized thermal plasma in
arbitrary geometry and apply it to describe the broadband X-ray continuum
of the galactic superaccreting microquasar SS433
observed by INTEGRAL.
A physical model of the X-ray emitting region is proposed that
includes thermal emission from the accretion disk, jets and hot corona where the photons of different origin are comptonized.  From comparison with INTEGRAL observations, we estimate physical parameters of
the complex X-ray emitting region in SS433 and present model spectra for
different viewing angles of the object.
\end{abstract}


\section{Introduction.}

   SS433 is a peculiar massive X-ray binary system with precessing
relativistic jets \citep{margon84}. The system is located at a
distance of about 5.2 kpc \citep{lockman&07} nearly in the
galactic plane. The optical companion V1343 Aql was first identified
in the survey of stars exhibiting $H_{\alpha}$ emission of
\citet{SS77}. It is one of the brightest stars in the Galaxy, the
bolometric luminosity of the object assuming isotropic radiation is
$L_{bol}\sim10^{40}$~erg/s \citep{Cherep82}. SS433 is a close binary
system with an orbital period of 13.1 days \citep{Cherep81}. The
uniqueness of this source comes from the presence of narrow oppositely
directed subrelativistic jets emitting red and blue-shifted
periodically variable lines. The commonly accepted model of SS433
suggests a continuous regime of supercritical accretion
of gas onto the relativistic star. The
optical star fills its critical Roche lobe supplying powerful and
almost continuous flow of gas onto the relativistic
star at a rate of $\sim10^{-4}M_{\odot}/yr$ \citep{Fabrika}. A
supercritical accretion disk forms together with narrow  jets of
gas propagating perpendicular to the disk plane from the central
regions of the disk and having a relativistic speed of 0.26c.

Presently, SS433 is recognized
as a galactic microquasar with precessing supercritical accretion disk
around a relativistic compact object, and has been extensively investigated
in the optical, radio and X-ray ranges (for a comprehensive
review and references see \citet{Fabrika}). Masses of both components were obtained from the analysis of the optical
light curves of SS433 \citep{AntCher87} and of the hard X-ray eclipses
observed by INTEGRAL (Cherepashchuk et al. 2008), so
the relativistic object in SS433 is most certainly a black hole.
Despite a significant
progress in understanding the nature of SS433, the mechanism of collimation
and acceleration of matter in jets to the relativistic velocity is
still unclear.

In the present paper we shall concentrate on the broad-band X-ray
spectrum of SS433. In the last years the source was intensively
observed by RXTE \citep{revnivtsev&04, filippova&06}, XMM
\citep{brinkmann&07}, Chandra \citep{marshall&05} and INTEGRAL
\citep{Cherep03,Cherep04,Cherep05,Cherep07}. INTEGRAL observations
of SS433 discovered a hard X-ray spectrum from the source up to 100
keV visible over the whole precessional period $\sim 162$ days. This
suggests the presence of an extended hot region in the central parts
of the supercritical accretion disk \citep{Cherep05,Cherep07,Cherep08}.

The initial analysis of the broadband X-ray continuum observed by
INTEGRAL \citep{Cherep05,Cherep07} allowed for the first time to
estimate the temperature and Thomson optical depth of the assumed
hot scattering corona above the accretion disk in SS433. However,
the spectral analysis made use of simplifying assumptions about
the corona geometry (mainly because of utilizing the standard
models incorporated in the XSPEC tool). The thermal emission from
the X-ray jet was treated independently and it was possible to
estimate the mass loss rate in the jet. The joint modeling
of hard X-ray eclipses and precessional variability allowed the
estimation of the size of hot corona in the framework of
the adopted geometrical model of the source
\citep{Cherep07,Cherep08}. It is well established from
observations (see \citet{Fabrika} for more detail) and
especially from recent studies of X-ray variability of SS433
\citep{revnivtsev&04, revnivtsev&06} that X-ray and optical
variabilities are correlated on time scales less than $10^7$~s.
The power spectrum analysis suggests that there is an extended
accretion disk with size similar to that in Cyg X-1 in the soft
state \citep{Titarchuk07}. The strong slow wind outflow from the
supercritical accretion disk forms a photosphere, and there is a
funnel filled with hot plasma around the jet. Clearly, to obtain
the correct physical picture, more accurate modeling of the
emission processes from the jet, accretion disk and hot corona is
needed.

\begin{figure}
\centerline{\hbox{\includegraphics[width=0.5\textwidth]{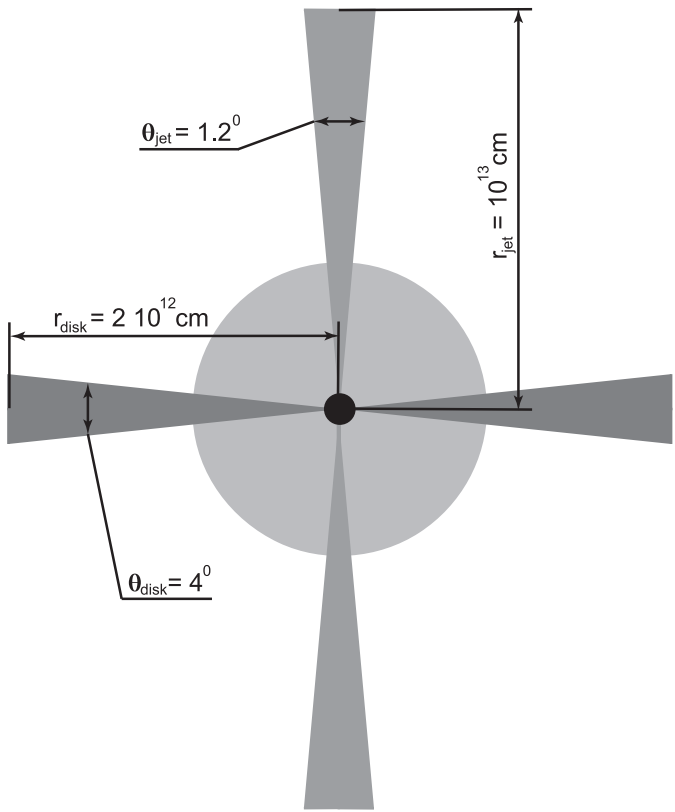}}}
\caption{Schematic picture of the model adopted for SS433}
\label{model}
\end{figure}

The purpose of this paper is to model the observed broad-band
X-ray continuum of SS433 using accurate treatment of emission
processes in the jet and the scattering hot corona above the
accretion disk. To calculate the broad-band X-ray spectrum, we
shall use the Monte-Carlo method (see the Appendix for a
more detailed description). The first reason for our choice is
connected with the geometry of the system containing the accretion
disk, the corona and the jet. This is a rather complicated
geometry, it has azimuthal symmetry, but for the outgoing
radiation nevertheless it is impossible to obtain any analytical
solutions, especially with account of the temperature decrease
along the jet. The second reason is connected with physical
processes, in particular the Compton scattering of thermal
photons from disk and jet by free electrons in the hot corona.
The accurate analytical treatment of this process in the radiation
transfer equation makes it impossible to solve it. For the same
reasons, the computational algorithm for numerical solution of the
transfer equation will be very sophisticated and will demand much
computational time. An advantage of the Monte-Carlo method is that
the computational algorithm is rather simple. It consists roughly
of one procedure repeated many times, and quite a small amount of
computational time is required to calculate the emergent spectra
with good accuracy. Moreover, when using finite-difference
methods, computational errors tend to increase with time as they
are accumulated at certain steps of the algorithm. In the
Monte-Carlo case the picture is completely different: the
precision of the result increases with computational time because
of larger number of individual tests and thus better statistics,
so waiting longer gives better result.

\section{Geometrical model of SS433}

The schematic picture of the source is presented in Fig.\ref{model}.
The collimated, oppositely directed jets were
proposed by \citet{fabrees79} and \citet{milgrom79} to explain the observed moving emission lines at unusual wavelengths.
These lines were interpreted as Doppler-shifted Balmer and He I lines \citep{margon&79}. The morphology of SS433 and the surrounding nebula from radio observations and the analysis of its temporal behavior provides strong support to
the idea of ballistically coasting matter ejected from the twin jets, which
justifies the so-called "kinematic model" of the jet precession \citep{abellmargon79} proposed to explain the periodic "movement" of the optical lines. The X-ray imaging observations of \citet{seward&80} also support the jet model.
The jets must be supplied with matter from some source, so a hypothesis of a compact star in a close binary system was proposed. The discovery of the 13-day periodicity in the radial velocity of the "stationary" SS433 emission-line system \citep{Crampton&80} proved this picture. The donor star fills its critical Roche lobe providing powerful flow onto the compact object. Thus, an accretion disk forms with a pair of narrow jets propagating perpendicular to the disk plane. The presence of the accretion disk follows also from the interpretation of the light curve that involves mutual eclipses of the massive early-type normal star and the extended accretion disk \citep{Cherep81,AntCher87}. As was already mentioned in the Introduction, the presence of a hot corona above the inner regions of the accretion disk was suggested to explain the hard X-ray spectrum of SS433 observed
by INTEGRAL over the entire precession period \citep{Cherep05, Cherep07, Cherep08}. So the considered model of SS433 including the accretion disk with hot corona above its central regions and two jets perpendicular to the disk plane is well justified by observations.

\subsection{Supercritical accretion disk}

In SS433, the optical star fills the Roche lobe and supplies matter
to the accretion disk at a rate of
$\sim10^{-4}M_{\odot}/yr$. The accretion
disk radius is assumed to be limited by
the size of the Roche lobe and thus for
$M_{bh}=7M_{\odot}$ and $q=M_{bh}/M_{opt}=0.3$ \citep{AntCher87} is
$\sim 10^{12}$~cm. The most emission from the disk is generated in
its central regions, so the exact value of the
accretion disk radius is of small importance. Far away from the central
parts, the standard accretion disk theory \citep{ShSyun73} can be
used to estimate its geometrical thickness. For the disk
height $h$ growing linearly with radius the disk opening
angle $\theta_{disk}$ can be directly found from the relation:
\begin{equation}
\label{diskheight} h=r\tan \frac{\theta_{disk}}{2},
\end{equation}
We shall consider an optically thick disk, so
in our calculations this conical surface
will be used as an opaque screen for incident photons.
For reasonable input parameters discussed below, the disk opening
angle is $\theta_{disk}=4^\circ$.

\subsection{Thermal X-ray jets}

The evidence for X-ray emitting jets in SS433 was found in
X-ray observations of the source. The \textit{EXOSAT} satellite
detected Doppler shifted iron lines moving periodically in
accordance with the precessional Doppler curve of the source
\citep{Watson_ea1986}, suggesting the X-ray emission origin from
the base of jets. Later observations by Ginga \citep{Kawai_ea1989,
Brinkmann_ea1991} and deep \textit{ASCA} observations
\citep{Kotani_ea1994, Kotani_ea1996} discovered many pairs of
Doppler-shifted emission lines originating in the blue and red
jets moving with a radial velocity of $0.26 c$ \citep{margon84}.
The soft X-ray emission is fitted by a combination of a thermal
bremsstrahlung continuum and emission lines from a
multi-temperature optically thin plasma in the conical jets with
length $L_j\sim 10^{13}$~cm and a total X-ray luminosity of $\sim
3\times 10^{35}$~erg/s \citep{Kotani1998}. More recent Chandra
\citep{marshall&05, Namiki_ea2003} and \textit{XMM} observations
\citep{Brinkmann_ea2005} confirmed the basic properties of X-ray
jets and put constraints on their geometrical and physical
parameters. The much higher kinetic luminosity of jets $\sim
10^{39}-10^{40}$~erg/s is derived from their interaction with the
surrounding nebula W50 (see \cite{Fabrika} for a detailed
discussion and further references). We shall assume X-ray jets to
have a conical shape, as inferred from the coincidence of the jet
opening angle $\theta_j$ in the optical and X-rays
\citep{Fabrika}, with the jet opening angle $\theta_j=1.2^\circ$.

\subsection{Hot corona}

The presence of a hot corona above the central parts of
the accretion disk in SS433 has been suggested by hard X-ray
observations of the source by \textit{INTEGRAL} \citep{Cherep05,
Cherep07, Cherep08}. The main indication is the unchanged from of
the hard X-ray spectrum at different precession phases. If only
thermal emission from cooling jets were responsible for the
observed X-rays, one would expect a significant softening of the
spectrum at the precessional phase $T3$ (disk edge-on), where the
innermost hottest parts of X-ray jets are screened by the
accretion disk edge. The observed primary X-ray eclipse light
curve and precessional hard X-ray variability cannot be
successfully described by thermal jets without a broad hot
structure above the central parts of the disk which is partially
eclipsed at the $T3$ precessional phase \citep{Cherep05,
Cherep08}. The hard X-ray spectrum at the maximum opening disk
phase can be modeled as scattered thermal radiation from jets by
an isothermal hot corona  ($T_{cor}\sim 18-30$~keV) with spherical
or slab geometry using different standard spectral tools from the
XSCPEC package (see, for example, \cite{Cherep07} for the ComPS
model fit). The corona around jets can be heated up to such
temperatures, for example, by the interaction of the precessing
jets with wind outflow from the supercritical accretion disk
\cite{Begelman_ea2006}. Motivated by additional geometrical
constraints obtained from the analysis of the joint precessional
and eclipsing X-ray variability of SS433 which suggests a fairly
broad opening angle of the jet funnel, we shall assume a spherical
shape of the corona. The parameters of the model are shown in
Fig.\ref{model}.

\section{Physical model of the X-ray emitting regions in SS433}

\begin{figure}
\centerline{\hbox{\includegraphics[width=0.5\textwidth]{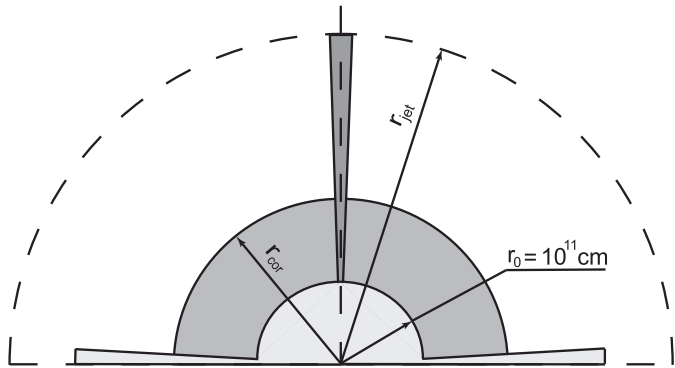}}}
\caption{The computational domain.}
\label{compdom}
\end{figure}

We consider the accretion disk with a spherical corona above its
inner regions, and two jets propagating perpendicular to the disk
plane. This picture is symmetric with respect to the disk plane,
so we shall consider only its upper half (Fig.\ref{model}). The system is
axially symmetric with respect to the jet axis, so radial profiles
of physical quantities fully characterize this 3-dimensional
structure. The computational domain (see Fig.\ref{compdom}) is bounded by the
surface of the accretion disk (from bottom) and a sphere with
radius equal to the jet height. The inner accretion disk is
assumed to have constant $h/r$ ratio, so the disk surface is
inclined by $2^\circ$ to the equatorial plane. The conical jet,
disk and corona begin at the coordinate origin. Both jet and
corona are assumed to consist of fully ionized hydrogen plasma.
Guided by observations of the accretion wind photosphere
\citep{Fabrika, revnivtsev&04}, we set the visible jet base at
$r_{0}=10^{11}$~cm and assume the inner spherical boundary of the
corona to lie at the same radius, as shown in Fig.\ref{compdom}.

\subsection{The jet}

We fix the jet opening angle at 1.2 degrees. The
temperature at the base of the jet is assumed to be the same as that of the corona and is set to be $T_0=19 keV$, the best-fit value found in our calculations.
\begin{equation}
\label{jettemperature} T=T_{0}\left(\frac{r_{0}}{r}\right)^{4/3},
T\sim \varrho^{2/3}, \varrho\sim r^{-2}
\end{equation}
In our simulations we shall assume that the jet temperature
changes with height according to the law corresponding to the jet
cooling due to adiabatic expansion of ideal monoatomic gas
(\ref{jettemperature}). In the paper of \citet{Kov&Sh} an
analytic solution for the jet temperature profile was obtained
taking into account both adiabatic and approximate treatment of
free-free emission cooling. For the constant jet opening angle
the solution obtained by Koval' \& Shakura is determined by the
mass rate in the jet, $\dot{M}_j$.  If   $\dot{M}_j\ge
\dot{M}_{lim}=(5\cdot10^{20})^{-1} \cdot 2\pi(1-\cos
\theta_{jet}/2){\cal R}\sqrt{T_{0}}(0.26c)^{2}r_{0}$, rapid
cooling of the jet plasma occurs. Here ${\cal R}$ is the universal
gas constant and $T_{0}=T_{cor}$ is the temperature at the base of
the jet. For jet parameters considered by us  $\dot{M}_{lim}\simeq
5.3\cdot10^{18}$~g/s. From observations, the kinetic luminosity of
the jet is about $L_{kin}\sim 10^{39-40}$~erg/s (\citet{Fabrika}),
so the lower limit for the mass loss rate in the jet is
$\dot{M}_j=3\cdot10^{19}$~g/s, which is six times greater than
$\dot{M}_{lim}$. For $\dot{M}_j/\dot{M}_{lim}<1$ the jet cooling
occurs almost as in the adiabatic case, but in the opposite case,
$\dot{M}/\dot{M}_{lim}>1$, the jet begins to cool rapidly with
radius and its total X-ray luminosity should be so small that it
would be impossible to observe it. The jet can have a larger
opening angle so that the value of $\dot{M}_{lim}$  increases and
may exceed $\dot{M}_j$, which will prevent the jet from rapid
cooling. There could be also some processes (Compton heating by
hard quanta from the corona, interaction with the wind from the
disk during the jet precession, etc.) that additionally heat the
jet and prevent it from rapid cooling. The accurate treatment of the jet heating
has not been done yet, so we assume the adiabatic
temperature distribution along the jet.

The density dependence along the jet follows from the
continuity equation. The jet mass outflow rate
$\dot{M_{j}}=nm_{p}VS$ , where $n$ is the number density of
protons at $r$, $m_{p}$ is the proton mass, $V=0.26c$ and $S=\pi
r^{2}\tan \theta_{jet}/2$ is the jet cross-section. So we can write
\begin{equation}
\label{density} n=n_{0}\left(\frac{r_{0}}{r}\right)^{2},
\end{equation}
where
\begin{equation}
\label{Dens0} n_{0}=\frac{\dot{M_{jet}}}{\pi m_{p} V r_{0}^{2}}
\tan^{2}\frac{\theta_{jet}}{2},
\end{equation}

\subsection{The corona}

The existence of hot coronas above highly luminous accretion disks
was first studied by \citet{BKB76, BKB77},
where two mechanisms responsible for corona formation have been considered. The first one is connected with disk radiation pressure. The particle moving in the vicinity of a radiating disk suffers the action of radiation from the whole disk. Thus, the radiation force depends on the position of the particle and is maximal at the central parts of the disk. When the disk luminosity approaches the value of $\sim 0.6$ Eddington limit, the radiation force exceeds the gravitational one and the outflow from the central parts of the disk sets in. The convection motions in the disk generate acoustic flux, which dissipates in the upper layers of the disk and heats them. As a result, a hot ($T\sim 10^9$~K) rarefied corona is formed above the central parts of the accretion disk. The second mechanism with account for magnetic field was studied by \citet{GRV79}. It is connected with magnetic field amplification in the disk. The magnetic field generated by differential rotation of conductive media is amplified due to convective motions in the direction perpendicular to the disk plane. The amplified field can attain strength comparable to equipartition value and magnetic flux will emerge from the disk, leading to formation of accretion disk corona, which consists of many magnetic loops, the fields of which provide an energy source for the corona heating. The corona is hot and has a low density compared to the accretion disk. In the present paper we assume
a spherical isothermal corona with temperature $19 keV =
2.2\cdot10^{8}$~K. For simplicity, the radial density dependence
in the corona is assumed to obey the same law as in the jet but
with the different initial value at $r=10^{11}$~cm. Then the
optical depth for Thomson scattering in the corona is
\begin{equation}
\label{taucor} \tau_{cor}=\sigma_{T}n_{0}\int
\limits_{r_{0}}^{r_{cor}} \frac{r_{0}^{2}}{r^{2}} \,dr
\end{equation}
The radius of the corona can be expressed in terms of
its optical depth as
\begin{eqnarray}
\label{rcor} & & r_{cor}=r_{0}/\left(\displaystyle 1-\frac{\tau_{cor}}{\sigma_{T}n_{cor}r_{0}}\right)\\
& & n_{cor}=(4.3-4.8)\cdot10^{12} cm^{-3} \nonumber\\
& & \sigma_{T}=6.65\cdot10^{-25} cm \nonumber
\end{eqnarray}
The analysis of the observed wide primary orbital hard X-ray
eclipse and the precessional variability \citep{Cherep07,Cherep08}
suggests large visible size of the hot corona comparable to that
of the accretion disk, but it is not absolutely certain. We have
fitted the observed X-ray spectra with two models of corona: 1)
the corona is not eclipsed by the precessing accretion
disk and has $r_{cor}=6.4\cdot10^{11}$~cm, $\tau_{cor}=0.24$; 2)
the corona is fully eclipsed with $r_{cor}=2.7\cdot10^{11}$~cm,
$\tau_{cor}=0.20$. Both models can reproduce the observed X-ray
spectrum, but the first model gives a slightly better fit. We
shall discuss  in more detail the obtained results below.


\subsection{The accretion disk}

Basic parameters of the standard accretion disk model include the
mass of the compact object $M$, mass accretion rate through the
disk $\dot{M}$, and the angular momentum transport efficiency
parameter $\alpha$. We are interested in the 'effective' mass
accretion rate through the disk and it is different from the mass
rate supplying by the optical star through the inner Lagrangian
point. The jet kinetic power is approximately equal to the
Eddington luminosity for the central black hole mass $M_{bh}=7
M_{\odot}$. This power corresponds to the critical accretion mass
rate through the disk $\dot M_{cr}$. As the same mass rate
should reach the central source to be ultimately consumed by the
black hole, the 'effective' mass flow rate through the disk is
$\dot{M}\approx 2\dot{M}_{cr}$. Taking $M=7 M_{\odot}$,
$\dot{M}\approx 2\dot{M}_{cr}$ and $\alpha=0.1$ we obtain that in
the outer gas-dominated region with free-free opacity
\citep{ShSyun73}, which lies beyond the radius
$r_{bc}=4\cdot10^{10}$~cm for the adopted parameters,
$$
h=6.1\cdot10^{3} \alpha^{-1/10}
\left(\frac{\dot{M}}{\dot{M_{cr}}}\right)^{3/20}
\left(\frac{M}{M_{Sun}}\right)^{9/10}
\left(\frac{R}{3R_{g}}\right)^{9/8}\\
$$
\begin{equation}
\label{heightShS} \qquad \left[1 -
\left(\frac{R}{3R_{g}}\right)^{-1/2}\right]^{3/20}
\end{equation}
At the accretion disk radius $R_{disk}=1.5\cdot10^{12}$~cm the
half-thickness of the disk is $5.5\cdot10^{10}$~cm, so the angle
between the disk surface and the equatorial plane in
(\ref{diskheight}) is $\theta_{disk}/2 \approx 2^\circ$ .
To calculate the emergent photon spectrum from the
corona, we shall assume that all accretion disk radiation is
emitted only in the central region, from a sphere with radius
$r_{0}$, while the outer regions of the accretion disk play the
role of an opaque screen for photons (Fig.\ref{compdom}).
Accretion disk photons are distributed uniformly across the sphere
with radius $r_{0}$ and initially move strictly in radial
direction.

\section{Results of simulations}

\subsection{Observational data}

\begin{figure}
\centerline{\hbox{\includegraphics[width=0.5\textwidth]{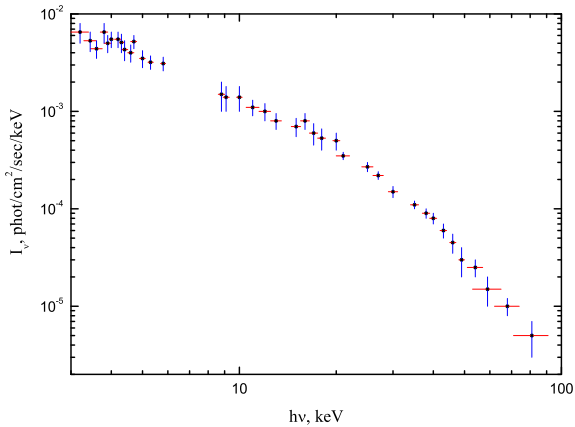}}}
\caption{X-ray continuum of SS433 from 3 to 90 keV}
\label{obs}
\end{figure}

For spectral analysis, we have made use INTEGRAL observations of
SS433 in May 2003 obtained simultaneously by IBIS/ISGRI (20-100
keV) and JEM-X (3-20 keV) telescopes \citep{Cherep03,Cherep05}.
The source was observed near the precessional phase 0 (the  T3
moment) when the accretion disk is maximally open and the flux
from the source is also maximal. The angle between the line of
sight and the blue jet (i.e. the jet pointing to the
observer) at this phase is $\approx 60^0$. The Doppler redshift
at this precessional phase is $z\simeq -0.1$. Spectral lines near
7~keV were ignored. The X-ray data were processed using the
standard OSA 5.1 INTEGRAL software developed by the INTEGRAL
science data center (ISDC, http://isdc.unige.ch \citep{ISDC}). The
resulting 3-90 keV X-ray spectrum is shown in Fig.\ref{obs}.

A model fit of the observed spectrum is accepted by achieving the
minimal reduced $\chi^2$ ($\chi^2_{red}=\chi^2/N$), $N=n-n_{par}$ is
the number of the degrees of freedom, see \citet{Zomb}.

\begin{figure}
\centerline{\hbox{\includegraphics[width=0.5\textwidth]{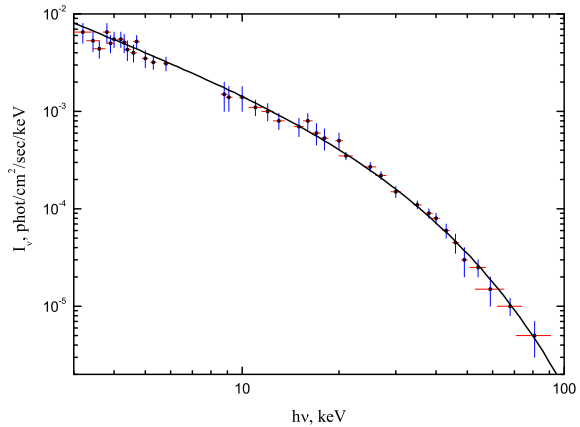}}}
\caption{The comparison of the Monte-Carlo simulated X-ray continuum of SS433 with observations.
$\theta_{obs}=60^{0}$, only the jet and corona emission are considered.
The radius of spherically-symmetric corona is $r_{cor}=6.4\cdot10^{11}$~cm,
the mass loss rate in the jet is $\dot{M_{jet}}=4\cdot10^{19}$~g/s. The viewing angle of
the system is  $\theta_{obs}=60^{0}$.}
\label{obssim6}
\end{figure}

\begin{figure}
\centerline{\hbox{\includegraphics[width=0.5\textwidth]{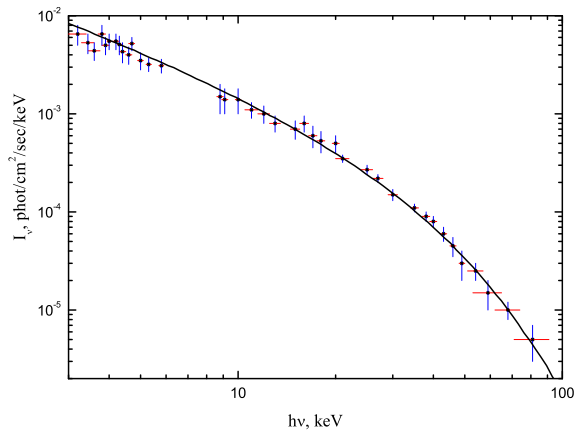}}}
\caption{The same as in Fig. \ref{obssim6} for smaller corona with
$r_{cor}=2.7\cdot10^{11}$~cm, and $\dot{M_{jet}}=3\cdot10^{19}$~g/s, viewed by
the same angle $\theta_{obs}=60^{0}$.}
\label{obssim3}
\end{figure}

\subsection{Bremsstrahlung comptonization model}

In this section we present the results of X-ray spectral modeling
neglecting the radiation from the disk. The distance to
SS433 is fixed at $d=5$~kpc $=1.5\cdot10^{24}$~cm. The emerging X-ray spectrum
in the 3-90 keV range is assumed to be formed due to free-free
emission of corona and jet with account for Compton scattering.
After many trials, the following best-fit parameters of the model
have been obtained (quantitative characteristics of the fitting are
discussed in the next paragraph). The temperature of the corona and
the jet base is $T_{cor}=19$~keV.
Fig.\ref{obssim6} shows the comparison of the
computational results with the observed broadband X-ray spectrum of
SS433 for a corona with $r_{cor}=6.4\cdot10^{11}$~cm,
$\tau_{cor}=0.24$, $n_{cor}=4.3\cdot10^{12} cm^{-3}$ which is not
fully eclipsed by the disk. The mass loss rate in the jet is $\dot{M}_j=
4\cdot10^{19}$~g/s, which corresponds to the kinetic luminosity of
$L_{kin}=1.2\cdot10^{39}$~erg/s. The next Fig.\ref{obssim3} shows the results
of simulations for a fully eclipsing corona with radius
$r_{cor}=2.7\cdot 10^{11}$~cm, the Thomson optical depth
$\tau_{cor}=0.2$, and the number density $n_{cor}=4.8\cdot10^{12} cm^{-3}$. The mass loss
rate in the jet is $\dot{M}_j= 3\cdot10^{19}$~g/s corresponding to the
jet kinetic luminosity $L_{kin}=9\cdot10^{38}$~erg/s.

It is clearly seen that in both cases
the model can describe the observed X-ray continuum.
The data includes 39 experimental points. There are 5 fitting parameters:
the jet mass loss rate $\dot{M}_j$, the coronal temperature $T_{cor}$,
number density at $r_0$, $n_{cor,0}$, the Thomson optical depth $\tau_{cor}$, and the initial radius $r_0$, so
the number of the degrees of freedom is $N_{dof}=34$. For
the model with non-eclipsing corona $\chi^2=21.4$ shown in Fig. \ref{obssim6}
the reduced $\chi^2_{red}=0.630$, and the confidence level of the fit at $N_{dof}=34$
is  $\approx 95.4\%$. For the second model with fully
eclipsing corona showmn in Fig. \ref{obssim3} $\chi^2=23.34$ for 34 degrees of
freedom, so the reduced $\chi^2_{red}=0.687$ and the
corresponding confidence level of the fit is $\approx 91.6\%$.
Thus the model with non-eclipsing corona (Fig.\ref{obssim6})
with smaller $\chi^2$ seems to be more realistic. All model parameters are listed in Table \ref{listpar}.

\begin{table}
\begin{tabular}{|l|c|c|}
\hline
Parameter & Eclipsing corona & Non-eclipsing corona\\
\hline
$r_{cor}, \cdot10^{11}$~cm & 2.7 & 6.4\\
$\tau_{cor}$ & 0.24 & 0.20 \\
$n_{cor}, \cdot10^{12} cm^{-3}$ & 4.8 & 4.3\\
$\dot{M}_j, \cdot10^{19}$~g/s & 3 & 4\\
$L_{kin}, \cdot10^{39}$~erg/s & 0.9 & 1.2\\
$\chi^2$ & 21.40 & 23.34\\
CL & $91.6\%$ & $95.4\%$\\
\hline
\end{tabular}
\label{listpar}
\caption{Physical parameters of two corona models}
\end{table}

\begin{figure}
\centerline{\hbox{\includegraphics[width=0.5\textwidth]{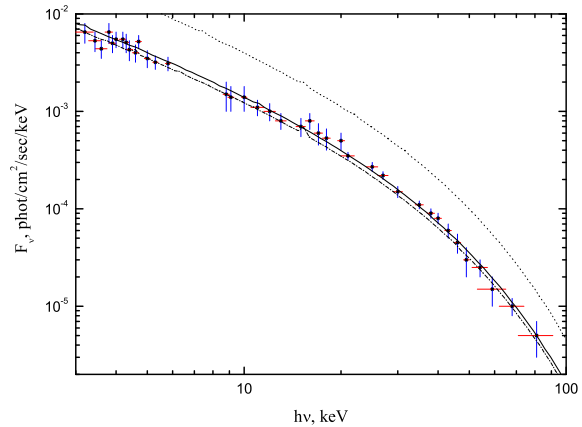}}}
\caption{The two-component model spectra of SS433 calculated for
different viewing angles: 60 degrees (the bold solid line), 82
degrees (the dash-dotted line), 90 degrees (the dotted line). The emergent
spectrum for 90 degrees was obtained by doubling the computational
data because of equal contribution of both jets and coronas.
$r_{cor}=6.4\cdot10^{11}$~cm, $\dot{M_{jet}}=4\cdot10^{19}$~g/s.}
\label{angdep}
\end{figure}

The angular dependence of the emergent X-ray spectrum is shown in
Fig.\ref{angdep}. The spectrum calculated for the viewing angle
$82^\circ$ is similar to $60^\circ$, the minimum viewing angle of
the jet during its precessional motion. The weak
dependence on the viewing angle may be due to the corona (which
radiates almost isotropically) mainly contributing to the total
emission at all angles. This figure also shows that the flux from
the system is at maximum for the viewing angle $90^\circ$ when
both jets and both parts of the corona contribute to the emergent
spectrum. The observations, however, suggest that the flux from
the source at this precessional phase (the moment T1, disk
edge-on) is minimal. This may indicate that the supercritical
accretion disk should be much thicker geometrically than predicted
by the standard model \citet{ShSyun73}. Significant screening of
the corona by a thick disk is needed to explain the observed
precessional variability of SS433 in hard X-rays
\citep{Cherep07,Cherep08}.

\begin{figure}
\centerline{\hbox{\includegraphics[width=0.5\textwidth]{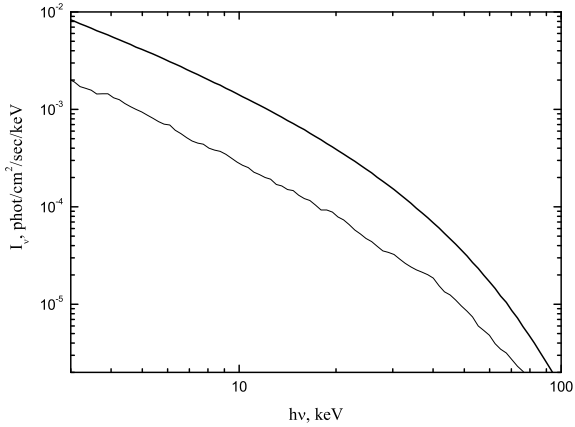}}}
\caption{The model spectrum of SS433 observed along the jet axis (the solid line) compared with spectrum at
$\theta_{obs}=60^{0}$ (the bold solid line).}
\label{00jet}
\end{figure}

The calculated spectrum of the source does not
vary significantly with the viewing angle in the allowed range $\theta_{obs}=60^\circ-90^\circ$,
but if we were looking straight along the jet ($\theta_{obs}=0^\circ$), we whould
observe a completely different picture (Fig.\ref{00jet}). \citet{FKAS, FAK} argued that ultraluminous
X-ray sources (ULX) are superaccretors like SS433 seen almost along the jet.
This hypothesis is based on the following assumptions: the most central
parts of the supercritical accretion disk are surrounded by
outflowing gas which is semi-opaque, so the
'edge-on' luminosity of such an object should be
much smaller than the 'face-on' luminosity, when
the hottest central regions are exposed directly.
We do not consider the
outflowing gas, which is not visible in observational data, so the
angular dependence of X-ray luminosity is completely different, as
demonstrated above. Moreover, if we were looking straight along the
jet, the luminosity would be actually minimal, since along the axis
the jet itself is optically thick for its own thermal emission.
The thermal energy power emitted per unit volume in the jet
decreases with distance as both
density and temperature decrease, so the most jet power
is released near the jet base. Thus, in order to see the hottest
central regions of the accretion disk, the system should be observed
not along the jet axis, but slightly from aside,
by viewing angles exceeding the jet opening angle.

\subsection{Influence of the accretion disk emission}

The two-component model described in the
previous section  gives satisfactory results, but
the accretion disk emission could also affect the
emergent X-ray continuum. To test this in our calculations,
we included the emission
from the inner parts of the accretion disk with
the following spectrum.

$$ \left\{
\begin{array}{rcl}
\label{fnu}
L_{\nu}&=&4L_{disk}/7\nu_{c}\left(\frac{\nu}{\nu_{c}}\right)^{1/3},  \nu<\nu_{c}\\
L_{\nu}&=&4L_{disk}/7\nu_{c}\exp\left(1-\frac{\nu}{\nu_{c}}\right),  \nu>\nu_{c}\\
\end{array}
\right.
$$
$$
\nu_{c}=\frac{hT_{ef}}{k}
$$
(Here $L_{disk}$ is the total luminosity of the accretion disk.)
$$
T_{ef} \equiv T_{ef}(r_{c})
$$
The value $r_{c}$, which is interpreted as the inner radius of
accretion disk, was taken to be $r_{sph}/0.3$, where
$r_{sph}=GM\dot{M}/L_{cr}$ is the spherization radius (see
\citet{ShSyun73}) at which the disk luminosity equals the critical
Eddington luminosity $L_{cr}=4\pi
cGM_{bh}/\kappa_{es}=8.8\cdot10^{38}$~erg/s. It was shown by
\citet{BKB77} that the outflow from a supercritical accretion disk
begins at $L_{d}=0.6L_{cr}$. Here we take the lower luminosity of
the disk $L_{d}=0.1L_{cr}$. This may be justified by analysis of
complicated physical processes in the supercritical accretion disk,
but the main reason for our choice is based on our simulations: at
higher $L_{d}$ an excess of soft X-rays appears, which is absent in
observations. For supercritical accretion the outflow can start
from radius where the luminosity is still subcritical, since the
disk should be additionally heated by radiation from hot corona. The
best fit with the observed spectrum occurs when the luminosity is
$0.1 L_{cr}$. Using the standard disk accretion equations and
Paczynski-Witta gravitational potential $\varphi_{PW}=GM/(r-r_{g})$,
the effective temperature can be found to be
\begin{equation}
\label{Tef}
T_{ef}(r)=2.05\cdot10^{7}\left(\frac{\dot{M}}{\dot{M}_{cr}}\frac{M_{sun}}{M_{bh}}\right)^{1/4}
\left(\frac{3r_{g}}{r}\right)^{3/4}\varphi^{1/4}_{r,r_{in}}
\end{equation}
Here $r_{in}=3r_{g}$ ,$r_{g}=2GM/c^{2}$ and $\varphi_{r,r_{in}}$ is
\citep{BK89}
$$
\varphi_{r,r_{in}}=\left(1-\sqrt{\frac{3r_{g}}{r}}\right)^{2}
\left(1-\frac{r_{g}}{3r}\right)\left(1+\frac{1}{2}\sqrt{\frac{3r_{g}}{r}}\right)\\
$$
\begin{equation}
\label{phir} \qquad \left(1-\frac{r_{g}}{r}\right)^{-3}
\end{equation}
For the assumed disk parameters we obtain from these formulas
$T_{ef}\approx 6.0\cdot10^{5}$~K $\approx 0.05$~keV.

\begin{figure}
\centerline{\hbox{\includegraphics[width=0.5\textwidth]{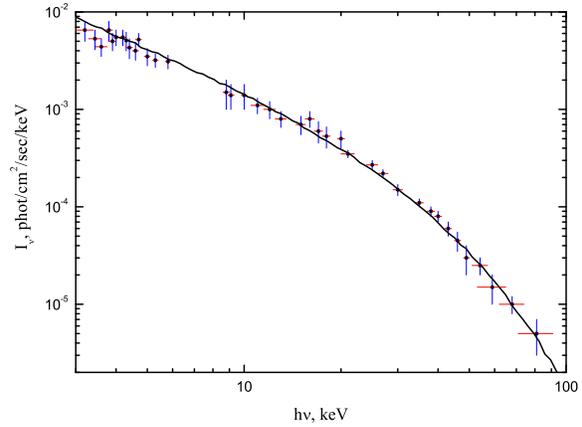}}}
\caption{The three-component model of SS443 spectrum. The model
spectrum is almost identical to the two-component
model shown in Fig.\ref{obssim3}. $r_{cor}=2.7\cdot10^{11}$~cm,
$\dot{M_{jet}}=3\cdot10^{19}$~g/s.}
\label{3cem}
\end{figure}

\begin{figure}
\centerline{\hbox{\includegraphics[width=0.5\textwidth]{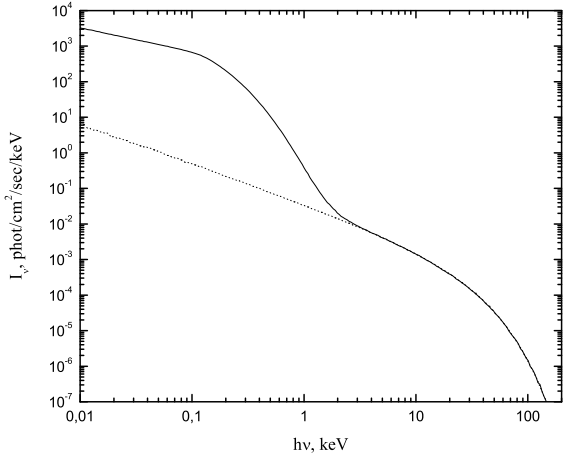}}}
\caption{Comparison of two- and three-component emission models. The
solid curve corresponds to the model with disk radiation,
the dashed one -- to the model without disk radiation.
$r_{cor}=6.4\cdot10^{11}$~cm, $\dot{M_{jet}}=4\cdot10^{19}$~g/s.}
\label{widerange}
\end{figure}

The accretion disk emission is assumed to be emerging from the sphere with
radius $r_0=10^{11}$~cm with all photons initially moving radially.
Inside this radius $r_{c}$ the heat production in the disk exceeds
the local Eddington value driving the mass outflow. We assume that
$r_{c}$ is approximately equal to the inner radius of the jet. We do
not take into account the radiation from the region inside $r_{c}$,
assuming that it is transformed into the jet energy and heating of
the corona.
The results of the three-component model simulations
are presented in
Fig.\ref{3cem} and \ref{widerange}.

It is clear that generally the disk radiation can strongly change the
emerging spectrum, but in the energy range 1-100 keV the spectral
difference is insignificant because of a comparatively low effective
temperature $T_{ef}\approx 0.05$~keV of the disk emission.
In principle, the spectrum at lower
energies can be used to estimate the accretion disk physical
parameters.

\begin{figure}
\centerline{\hbox{\includegraphics[width=0.5\textwidth]{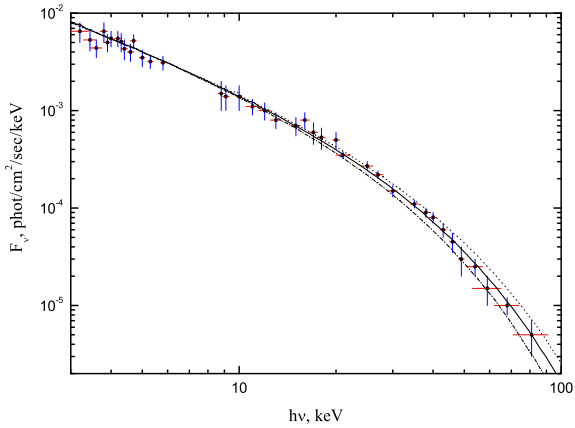}}}
\caption{The results of the simulations for different corona temperatures: 17 keV (the dash-dotted line), 19 keV (the solid line), 21 keV (the dotted line). The best fit is found for $T_{cor}=19$~keV,  $\tau_{cor}=0.24$, $r_{cor}=6.4\cdot10^{11}$~cm, $\dot{M_{jet}}=4\cdot10^{19}$~g/s.}
\label{fitcorT}
\end{figure}

\begin{figure}
\centerline{\hbox{\includegraphics[width=0.5\textwidth]{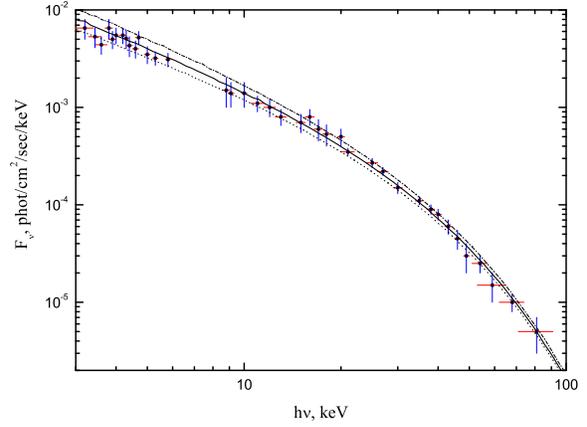}}}
\caption{The results of the simulations for different values of
the jet mass loss rate: $3\cdot10^{19}$~g/s (the dotted line), $4\cdot10^{19}$~g/s (the solid line), $5\cdot10^{19}$~g/s (the dash-dotted line). The best fit is found for  $\dot{M_{jet}}=4\cdot10^{19}$~g/s, $\tau_{cor}=0.24$, $T_{cor}=19$~keV, $r_{cor}=6.4\cdot10^{11}$~cm.}
\label{fitjet}
\end{figure}

\begin{figure}
\centerline{\hbox{\includegraphics[width=0.5\textwidth]{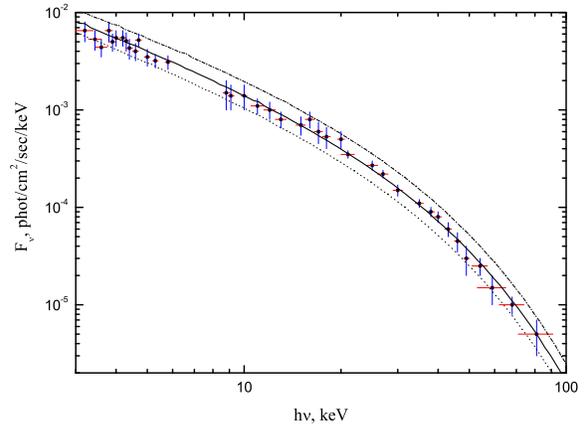}}}
\caption{The results of the simulations for different optical depth
of the corona: 0.2 (the dotted line), 0.24 (the solid line),  0.3 (the dash-dotted line). The best fit is found for $\tau_{cor}=0.24$, $T_{cor}=19$~keV, $r_{cor}=6.4\cdot10^{11}$~cm, $\dot{M_{jet}}=4\cdot10^{19}$~g/s.}
\label{fitcortau}
\end{figure}

\begin{figure}
\centerline{\hbox{\includegraphics[width=0.5\textwidth]{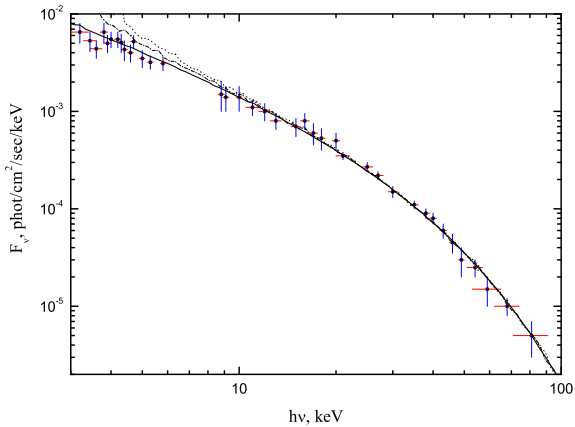}}}
\caption{The results of the simulations for different values of the  accretion disk effective temperature: $6.0\cdot10^{5}$~K (the solid line), $9.4\cdot10^{5}$~K (the dash-dotted line), $1.1\cdot10^{6}$~K (the dotted line). The best fit is found for $T_{ef}=6.0\cdot10^{5}$~K, $\tau_{cor}=0.24$, $T_{cor}=19$~keV, $r_{cor}=6.4\cdot10^{11}$~cm, $\dot{M_{jet}}=4\cdot10^{19}$~g/s.}
\label{fitdisk}
\end{figure}

\begin{figure}
\centerline{\hbox{\includegraphics[width=0.5\textwidth]{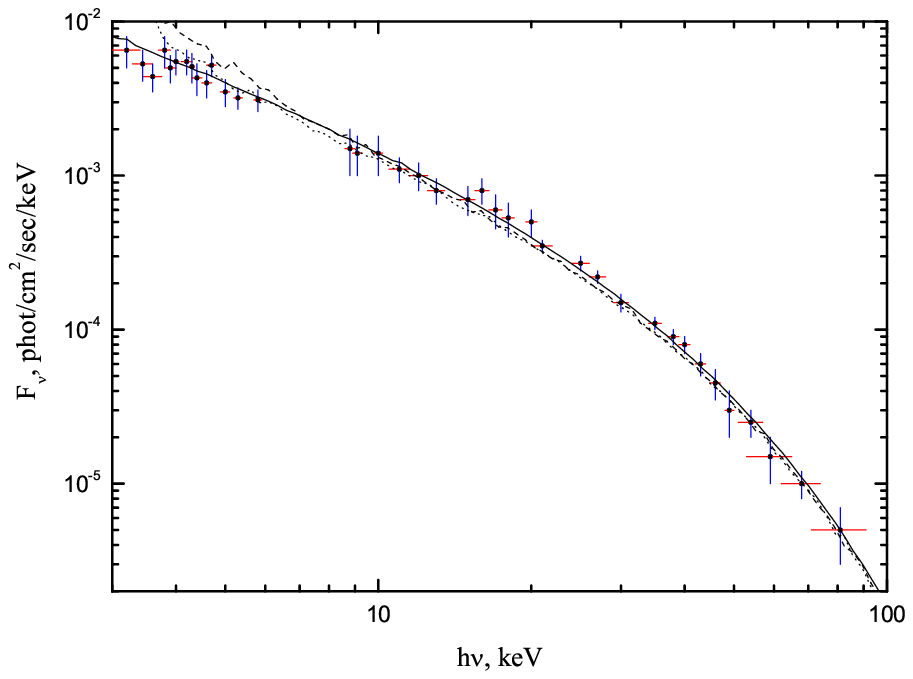}}}
\caption{Fitting the observational data for a model with low kinetic luminosity of the jet for different values of the accretion disk effective temperature: $9.4\cdot10^{5}$~K (the dotted line), $1.1\cdot10^{6}$~K (the dashed line).  $\tau_{cor}=0.24$, $T_{cor}=19$~keV, $r_{cor}=6.4\cdot10^{11}$~cm, $\dot{M_{jet}}=3\cdot10^{19}$~g/s. The solid line corresponds to $T_{ef}=6.0\cdot10^{5}$~K, $\dot{M_{jet}}=4\cdot10^{19}$~g/s with all other parameters as shown for previous lines.}
\label{fitjetdisk}
\end{figure}

\section{Discussion}

We have developed a Monte-Carlo code for modeling X-ray  (3-90
keV) continuum of SS433 based on the realistic physical model of
the source, including the accretion disk, thermal jet and hot
scattering corona. We found the model parameters that
provide good qualitative and quantitative agreement with INTEGRAL
observations. The fitting procedure was as follows.

The similarity of hard X-ray spectra of SS433 at different
phases of the precessional period, as well as the joint modeling
of the orbital primary X-ray eclipse and precessional variability,
suggest the presence of a hot corona that is not fully eclipsed by
the donor star \citep{Cherep08}. This fact allows us to put the
lower limit on the outer radius of the corona.

First, we adjust the shape of the spectrum in hard X-rays by varying the coronal temperature,
as shown in Fig.\ref{fitcorT}. It is seen from this figure that the coronal temperature
affects the hard X-ray spectrum, but does not change the soft X-ray continuum.
In opposite, the change of $\dot{M_{jet}}$ affects the form of only the soft X-ray
spectrum (Fig.\ref{fitjet}) and the hard X-ray is left unchanged, so we can
infer the jet mass-loss rate $\dot{M_{jet}}$ by adjusting the spectrum in the soft
X-rays (the temperature of the jet base is assumed to be equal to that of the corona). Note that the change of both jet and corona parameters cannot compensate
each other because the related spectral components affect the resulting
spectrum in different energy ranges. After that, we adjust the corona optical depth
(Fig.\ref{fitcortau}). The variations of coronal
optical depth at $\tau_{cor}<1$ do not change the form of the spectrum,
but do change the coronal X-ray luminosity. So we fit
the luminosity of the coronal component as shown in Fig.\ref{fitcortau}.
Finally, we constraint the accretion disk emission temperature
at which the soft part of the model spectrum is still consistent
with observations (Fig.\ref{fitdisk}). When the effective temperature
of the accretion disk emission is too high,
a soft X-ray excess appears, which is not present in observations.
In Fig.\ref{fitjetdisk} we present the results of our simulations
when the X-ray excess from the accretion disk partially compensates
the low energetics of the jet. This figure  clearly demonstrates
that these model parameters do not give quantitative agreement with
observations.

The last three figures clearly show that the emerging
X-ray continuum is very sensitive to slight changes
of physical parameters and our results form an island
in the parameter space. Further observations and more precise
determination of the distance to the source, the component mass ratio,
etc., can change the jet mass loss rate and corona density.

We also should stress that the hard X-ray spectrum of
SS433 by itself can be formally fitted in several physical models,
ranging from purely thermal emission from jets with changing
temperature to variants of standard comptonization procedures from
the XSPEC package. An example of such a description with CompPS
model is given in \cite{Cherep07}. However, when the physical
parameters of the system are fixed by other observations, like
those discussed in this paper, the description of hard X-ray
spectrum by standard tools become not self-consistent. So the
results of spectral modeling presented in this paper can be
considered as a first attempt to find disk, jet and corona
parameters in a self-consistent way. Clearly, further observations
of the complicated variable emission from SS433 are needed to
improve the model parameters.

We would like to draw attention to some recent results in
determining the binary component masses in SS433. The optical
spectroscopy of the system \citep{hillwig08, hillwig04} revealed
the presence of absorption lines in the spectrum of the optical
component identified as a $\sim$~A7I supergiant star. Observed
orbital Doppler shifts of the absorption lines of the optical
component allowed Hillwig \& Gies to determine the mass ratio of
the relativistic ($M_x$) and the optical ($M_v$) components in
SS433 $q=M_x/M_v\simeq 0.3\pm0.11$, implying the masses
$M_x=4.3\pm0.8 M_\odot$ and $M_v=12.3\pm 3.3 M_\odot$. These
results are not the final ones (cf. the analysis by
Cherepashchuk et al. (2008), who obtained $M_x\sim 5 M_\odot$),
and the values of the binary component masses can change with
different interpretation of optical spectroscopic observations, as
those authors mention in the conclusion \citet{hillwig08}.
Nevertheless, the smaller mass of the compact object compared to
our choice ($M_{BH}=7 M_\odot$) will result in decreasing of the
corresponding Roche lobe size, and thus the accretion disk radius.
But X-ray spectral properties of SS433 are determined by jets and
corona above central parts of the accretion disk, so the X-ray
spectrum of the source will hardly change significantly for
smaller $M_x$.

We conclude that
the observed broadband X-ray continuum of SS433 is dominated by thermal
emission of the jet and rarefied hot corona around the inner parts of
the supercritical accretion disk, with addition from the inverse
Compton scattering on hot electrons of the corona. The jet mostly
contributes in the soft X-ray band, while hard X-ray emission is
dominated by thermal emission from the corona and inverse Compton
scattering of soft thermal X-ray photons. The emission from inner
regions of the accretion disk also contributes to the spectrum, but
mostly at energies below 1~keV. The observed broadband X-ray spectrum of SS433
is best fitted by the corona temperature $\simeq 19$~keV with
Thomson optical depth $\tau_{cor}\simeq 0.24$ and the jet mass loss
rate about $4\cdot10^{19}$~g/s which corresponds to the kinetic
luminosity of the source $1.2\cdot10^{39}$~erg/s.

\section{Acknowledgements}

The work is partially supported by the
Russian Foundation for Basic Research under grants 08-02-08494, 08-02-00491, 06-02-90864, 06-02-91157,
by the Russian Federation President's grant for supporting of leading scientific schools NSh-2977.2008.2
and by the RAS Presidium program.

The authors would like to thank A.G. Doroshkevich and B.E. Stern for useful discussion
and the anonymous referee whose remarks helped us to improve the paper.

\section{Appendix: Monte-Carlo Techniques}

The Monte-Carlo method has another name: the statistical test
method. An innumerable number of processes can be listed the
outcome of which is not determined and has a probabilistic nature.
When repeated, they will give us different results. But these
processes obey statistical laws that can be ascertained by
repeating one process many times, i.e. by performing statistical
tests. As the number of tests increases, the fluctuations decrease,
and we can evaluate some value that we are interested in. For
instance, the mean value of a random variable, its variance,
the density probability, etc. The process of photon propagation
through a medium has probabilistic character: the photon interacts
with matter, this interaction changes the photon properties (e.g.
its frequency, the direction of motion) and one cannot predict exactly
the outcome of the process, i.e. whether the photon escapes from the
medium, where the photon will escape, and its new frequency.
Because of that the Monte-Carlo method is suitable for deriving
the characteristics of the radiation field, such as its spatial
and frequency distribution. Of course, one can solve the radiation
transfer equation to obtain these quantities
using some approximations for stationary and time-dependent problems,
e.g. see \citet{SunTit80}, \citet{Titarchuk88}, but in all cases
the geometry of the problem is rather simple and in many cases cannot be applied directly
to sources with complec geometry of emitting/scattering regions,
like SS433. Also, the finite-difference numerical solution of an integro-differential equation
with frequency dependence is quite sophisticated and requires much
computational time. These are our reasons for choosing the Monte-Carlo method.

One possible way to perform a single test with a photon is to simulate
its trajectory in a medium. Initially a photon is placed somewhere
within the computational domain, its frequency and the direction of
motion are known, and we trace the photon as it propagates through
the medium (below we shall discuss how one can do this). The moment
when the photon leaves the computational domain, or its trajectory
is terminated, is considered to be the end of the test and the
longest time of one test is taken to be the duration of the
Monte-Carlo experiment itself $\Delta t$. But there is a problem:
any real source emits enormous amount of photons, so it will take
correspondingly enormous amount of time to simulate all trajectories,
and one will have to wait for years to get the result. It is clear
that we should abandon real photons and introduce model ones. One
possible way is to trace many photons combined in a group, i.e. a photon packet.
At this point, another problem arises: when a pack of photons
interacts with the medium, some photons are, for example, scattered,
others are absorbed or produce an electron-positron pair in the
field of an atomic nucleus, or something else. One can split the
photon packet and follow the trajectory of each part, but there is
another way out. In our simulations we used the indivisible photon
packets method developed by \citet{Lucy L.B.}. According to this
method, individual photons are grouped in the packets with constant
energy
\begin{equation}
\label{eps0} \varepsilon_{0}=nh\nu.
\end{equation}
Such a trick helps us to avoid individual tracing of a large number of
photons, especially in the low-energy part of the spectrum, since
the number of photons in a packet increases with decreasing
frequency. The process of interaction of a photon from the packet
is chosen each time, according to the known probability of the variant.
Such a method may look
inexact, but the increasing of the number of photon packets makes less
probable things happen, which approaches the result to the exact
solution.

\subsection{The number of photon packets}

\begin{figure}
\centerline{\hbox{\includegraphics[width=0.5\textwidth]{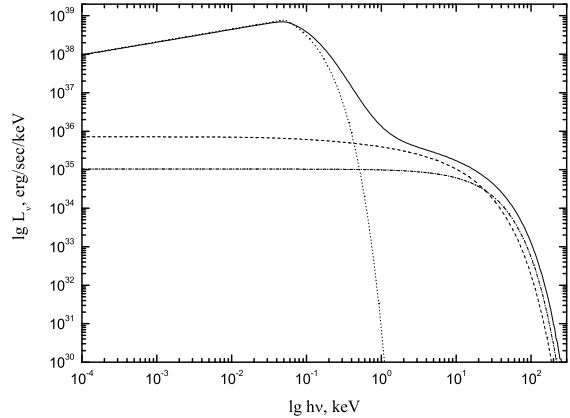}}}
\caption{Initial spectra: accretion disk emission (dotted line), jet
free-free emission (dashed line), corona free-free emission (dash-dotted line). The solid line is the result of the Monte-Carlo simulation. Comptonization of emission on corona electrons is responsible for 'hardening' of the spectrum.}
\label{inspec}
\end{figure}

At the beginning of the Monte-Carlo experiment a number of the
photon packets is placed within the computational domain. This
number is arbitrary, but the higher this number, the better accuracy
can be achieved in the resulting spectrum. As a consequence of the
statistical character of the experiment, the relative error in the
result is inversely proportional to the square root of the photon
packets number, so to increase the accuracy  by an
order of magnitude one should increase the number of photon packets
by two orders of magnitude. About $10^{8}$ photon packets is enough
to provide a smooth spectral curve that can be compared with
observations. In our case there are photon packets of different
origin: free-free photons from  the jet and corona and the
accretion disk photons. Each 'photon type' has its own area of the initial spatial location and spectral
distribution. In Fig.\ref{inspec} the initial spectra are presented together with the result of the Monte-Carlo simulation
which is discussed in the main text. Each component makes its contribution to the source luminosity.
The free-free emissivity of plasma in the jet or corona is
\begin{equation}
\label{Jnu} L_{ff}=\int \,dV n_{e}^{2}(r) \int \,d\Omega \int
\limits_{0}^{\infty} \,d\nu k_{\nu}(T)B_{\nu}(T)
\end{equation}

It should be taken into account that the temperature varies along
the jet, so one should first integrate over the
frequency and then over the jet volume. The corona is isothermal
and $L_{ff,cor}$ can be calculated as a product of three integrals.
The photon luminosity of the accretion disk was taken to be
a fraction of the Eddington luminosity $L_{disk}=\eta L_{Edd}$
with the factor $\eta$ being inferred during the fitting procedure.

One should set the number of photon packets $N$ that will
participate in the Monte-Carlo experiment. After calculating the
total luminosity of all components
\begin{equation}
\label{Ltot} L_{tot}=L_{disk}+L_{ff,jet}+L_{ff,cor},
\end{equation}
we determine the value $\varepsilon_{0}/\Delta t$ as
\begin{equation}
\label{eps0dt} \frac{\varepsilon_{0}}{\Delta t}=\frac{L_{tot}}{N},
\end{equation}
where $\varepsilon_{0}$ is the energy of the packet, $\Delta t$ is
the duration of the Monte-Carlo experiment. Once this value is
calculated we can determine the number of photon packets
representing different spectral components as
\begin{equation}
\label{Ntoti} N_{i}=\frac{L_{i}}{\varepsilon_{0}/\Delta
t}=\frac{L_{i}}{L_{tot}}N.
\end{equation}
The index 'i' stands for the 'type of emission' (corona or jet free-free
or accretion disk emission).

\subsection{The distribution of photon packets over the frequency bins}

The frequency to the photon packet should be assigned. The necessary
spectral range is divided into several frequency bins. Their number must
be sufficiently high to provide a smooth spectrum, but small
enough so that the average number of photons (packets) per bin
$N_{ph}$ be larger than unity, $N_{ph}/N_{bin}\gg1$. In our
calculations we divided the spectrum into 500 equal bins on
logarithmic scale. The initial frequency was taken in the middle
of the frequency bin. The number of photon packets in each
frequency bin $N_{ij}$ is determined by the photon spectral
distribution and is proportional to the contribution of the
frequency bin to the emissivity of the spectral component:
\begin{equation}
\label{NiNtot} N_{ij}/N_{i}=\int \limits_{\nu_{j-1}}^{\nu_{j}}
L_{i,\nu} \,d\nu/\int \limits_0^\infty L_{i,\nu} \,d\nu
\end{equation}
All photon packets in each frequency bin are assigned the
same frequency. During a Monte-Carlo experiment, the trial photons
are chosen sequentially starting from the low-frequency bins to
the high-frequency ones.

\subsection{Initial data for photon packets}

\subsubsection{The spatial position}

Let us determine the spatial positions of the photon packets.
Obviously, in a homogeneous and isothermal medium the
free-free photons should be distributed uniformly. But if the medium
is inhomogeneous or/and non-isothermal, the situation is more
complicated. One should divide the whole volume into regions
of constant density and temperature. In the particular case of
spherically symmetrical density and temperature distributions in the
corona, the volume was divided into spherical layers with constant density and
temperature. The corona was divided into 300 layers of constant
thickness, and jet was divided by similar spherical cuts into about
$10^4$  truncated cone layers. The contribution of each layer
to the total emissivity was calculated according to (\ref{Jnu}) and a certain
number of photon packets representing the free-free spectrum of
each layer was introduced. In each layer its own free-free spectrum was
maintained constant. The photon packets within a layer were distributed
with equal probability for a free-free photon
to appear in each point in the layer
$$
\left\{
\begin{array}{rcl}
\label{incoff}
r&=&\left[\left(r_{i}^{3}-r_{i-1}^{3}\right)\gamma_{1}+r_{i-1}^{3}\right]^{\frac{1}{3}}\\
\cos\theta_{in}&=&\left( \cos \theta_{jet}/2-\sin \theta_{disk}/2
\right)\gamma_{2}+\sin \theta_{disk}/2\\
\varphi_{in}&=&2\pi\gamma_{3}
\end{array}
\right.
$$
Here $r_{i}$ and $r_{i-1}$ are outer radii of the $i$-th layer,
$\gamma_{1,2,3}$ are random numbers distributed uniformly over the
interval $(0,1)$, $\theta_{jet}$ is the jet opening angle
and $\theta_{disk}$ is the accretion disk thickness angle.

In the case of the accretion disk photon packets, the situation does
not differ drastically. They are emitted from the spherical segment
with radius $r_{0}$, and should be uniformly distributed over this
surface:
$$
\left\{
\begin{array}{rcl}
\label{incod}
r&=&r_{0}\\
\cos\theta_{in}&=&\left(1-\sin \theta_{disk}/2 \right)\gamma_{2}+\sin \theta_{disk}/2\\
\varphi_{in}&=&2 \pi \gamma_{3}
\end{array}
\right.
$$

\subsubsection{The initial direction of motion}

One should set the initial direction of motion for the photon
packets. Free-free photon packets emitted in a plasma move isotropically \citep{Sobol'73}.
$$
\left\{
\begin{array}{rcl}
\label{initdir}
\cos\theta&=&2\gamma_{4}-1\\
\varphi&=&2\pi\gamma_{5}\\
\end{array}
\right.
$$
In the Cartesian coordinates, unit vectors (orts) determining the initial
direction of motion are written as
$$
\left\{
\begin{array}{rcl}
\label{invectff}
e_{x}&=&\sin\theta\cos\varphi\\
e_{y}&=&\sin\theta\sin\varphi\\
e_{z}&=&\cos\theta\\
\end{array}
\right.
$$
Initial photon packets from the accretion disk are assumed to
move radially.

\subsection{Simulation of the photon packet trajectory}

Simulation of propagation of photons through the medium is the
essence of the Monte-Carlo experiment because at this stage the
resulting angle and spectral distribution of the photon packets
is formed. Once the initial conditions for a photon
packet are set, one can find a randomly distributed optical depth the
packet must travel before it interacts with matter using
a simple Monte-Carlo formula
\begin{equation}
\label{taur} \tau=-\ln\gamma, \qquad \gamma\quad {\rm is \,\,\,
a\,\,\, random\,\,\, number.}
\end{equation}

Next, we should choose the velocity and the direction of motion of
an electron that scatters the photon packet. For electrons with
relativistic Maxwellian distribution, the velocity of electron can
be calculated using the random number algorithm suggested by
\citet{PSS83}. In the corona all directions of motion are equally
probable, so the distribution of orts of the electron motion is
isotropic (\ref{initdir}).
The matter in the jet moves with the radial velocity $0.26c$. The distribution of
directions of motion of electrons is isotropic in the local
co-moving jet frame. So the resulting vector of motion of an electron
is obtained by the Lorentz transformation of the isotropic
ort (in the co-moving frame) to the rest frame.

After that the coordinates of the photon-electron
interaction point can be found. We determine the total optical depth as
\begin{equation}
\label{tautr}
\tau=\int\,dln_{e}(r)\left[\sigma_{C}+k_{\nu}(T)n_{e}(r)\right]
\end{equation}
For a given density distribution we can integrate this
expression analytically, taking into account that the photon
packet trajectory is a straight line defined by the
equation
\begin{equation}
\label{ri} \vec{r}_{i+1}=\vec{r}_{i}+s\vec{e}_{v}
\end{equation}
Thus we obtain a transcendental equation for the distance $s$
the photon travels between two subsequent interactions. This equation
is solved numerically using the Newton method.
After that we obtain the
required coordinates of the interaction point from (\ref{ri}). We
must check whether the interaction point lies within the region occupied
by plasma (the computational domain). If it does not, then we trace the
photon trajectory to the boundary of the computational domain, and
collect all necessary information, such as photon's frequency and escape
angle.

If the interaction point lies within the computational domain, we
should choose the process of interaction of radiation with
matter. We assume that the continuum X-ray spectrum of SS433
in the 3-90 keV range is formed due to the Compton scattering by
free electrons and free-free absorption/emission, so the choice
should be done randomly between these two events. We select the
physical event by means of a simple Monte-Carlo procedure. A random
number uniformly distributed over the interval $(0,1)$ is selected,
and if it satisfies the criterion \citep{Lucy L.B.}
\begin{equation}
\label{choose}
\gamma<\frac{\sigma_{C}}{\sigma_{C}+k_{\nu}(T)n_{e}(r)}
\end{equation}
then the photon packet was considered to suffer a Compton
scattering, otherwise it was considered to suffer a free-free
absorption. If scattering occurs, we should determine the new
frequency of the packet and its new direction of motion as in
\citet{PSS77}. The energy of the photon packet should be changed
with the change of frequency according to (\ref{eps0}),
so the number of photons in the packet remains constant.
If the photon packet is absorbed, then it simply
disappears, and we proceed to the next packet.

This sequence should be repeated until the photon packet
is absorbed or has left the computational domain.
Then we gather all required information about the escaped photon packet.
To complete the Monte-Carlo experiment,
this procedure is applied to all photon packets.

\bsp

\label{lastpage}

\end{document}